# An Active Matter Pathway for Non-Equilibrium Liquid-Liquid Extractions

Jude C. Obijiaku, Oladipupo P. Ogolo, Damilola E. Fadipe, Angela Deneke, and Karthik Nayani*

Ralph E. Martin Department of Chemical Engineering, University of Arkansas, Fayetteville, AR, 72701, USA

We report on a Marangoni flow-driven pathway for non-equilibrium transport of metal ions across an oil-water interface. Under specific conditions, we show that the adsorption of a metallic species onto the extractant-decorated oil-water interface creates interfacial tension gradients that give rise to Marangoni flows. Strikingly, we observe that these flows are accompanied by removal of metal from the aqueous phase and that the extraction kinetics follow a non-diffusive behavior. The extraction rates are significantly higher - at least a threefold increase - when Marangoni flows are present. We show that the strength of the interfacial instability is coupled to the lowering of the interfacial tension- which is metal-dependent and therefore can lead to rate-mediated selective extractions. Overall, our study shows a fundamentally different paradigm for liquid-liquid extractions opening up a range of new directions of inquiry and future technologies.

## 1. Introduction

In chemical separations, equilibrium mass transfer plays a critical role in transport of the species of interest [1]. For instance, in solvent-mediated extractions, the difference in chemical potential of the solute between two immiscible phases is the driving force for transport and diffusion plays a key role in distributing the solute between the phases until equilibrium is reached [1, 2]. The efficiency and extraction rate depend strongly on the rate at which these molecules are transported across the interface- often controlled by interfacial resistance, diffusion within boundary layers, and the stability of the interface itself [2-7]. Under equilibrium conditions, the transport rate is therefore constrained by diffusion-limited molecular exchange and slow interfacial renewal [1, 7-11]. Hence, there is increasing interest in the development and understanding of non-equilibrium pathways that could promote faster mixing, transient concentration gradients, and dynamic restructuring of the interface, all of which increase the rate of molecular transfer [1, 12-16].

In systems with liquid-liquid interfaces, fluctuations in temperatures or surfactant concentrations result in local changes to the interfacial tensions. The gradients of the interfacial tension result in an instability that reveals itself as spontaneously arising interfacial convection which propagates into the bulk due to viscous forces[17-26]. In recent times, particular attention has been devoted to droplets, that propel themselves via unbalanced tangential stress gradients arising from interfacial flows[27-33]. The broad field exploring these far from equilibrium phenomena being termed as “active matter”.

Here, we introduce an active matter-based approach for non-equilibrium transport of metal cations across an oil-water interface. Active matter systems self-propel via mechanisms that continuously consume energy to generate directed motion [34-36]. This sustained energy input maintains the system far from equilibrium, and holds the potential to enable new transport pathways in the context of liquid-liquid extractions that may overcome mass transfer limitations [2, 10, 37, 38]. We describe phenomenology pertaining to a commonly used extractant di(2-ethylhexyl) phosphoric acid (HDEHP or DEHPA, molecular structure in Figure 1A). HDEHP when added to an oil-water interface[24, 26, 39-42] (under specific conditions), resulted in strong Marangoni flows which stem from gradients in the interfacial activity/tension[12, 34, 35, 43]. These flows eventuated in a subsequent extraction of metals[13-15, 44-50] initially dispersed in the aqueous phase. Strikingly, we observe that: 1) extraction kinetics follow a non-diffusive extraction behavior when the Marangoni flow is present. 2) Interfacial tensions and velocity of flow are sensitive to metal-surfactant interactions, thereby enabling a pathway for selective extractions. Therefore, by integrating these fields (active matter and solvent extraction), this study provides insights into harnessing liquid-liquid interfacial instability-driven pathways for separation methods.

[a] *Department of Chemical Engineering, University of Arkansas, 3202 Bell Engineering Centre, Fayetteville, AR 72701, USA*
* *Email: knayani@uark.edu*

## 2. Results and Discussion

**2.1 Molecular hypothesis**: Figure 1A illustrates the pH-responsive protonation states of HDEHP (pKa = 3.24)[13, 43] . At pH~ 3, HDEHP exists predominantly in its protonated form, whereas at pH~7, HDEHP is largely deprotonated [43]. The molecular hypothesis driving this work is presented in Figures 1B-E. At pH ~3, HDEHP confined to a water-oil interface is protonated and therefore, no interfacial tension gradients exist (as presented in Figure 1B). Similarly, at pH~ 7, HDEHP at the interface is completely deprotonated and there exists no gradients of interfacial tension in this scenario either (Figure 1C). Figure 1D presents a scenario wherein the aqueous phase at pH ~7 has a metal cation ($Zn^{2+}$) with strong affinity for HDEHP solvated within it [51-54]. In this instance, the surfactant-metal bound complexes, and the deprotonated surfactant (without bound metal cations) likely have different surface activity. This results in the possibility of interfacial tension gradients. Figure 1E, depicts the sustenance of the instability via the transport of the bound metal-surfactant complex into the oil phase and the replenishment of free surfactant at the interface (if the concentration of HDEHP in the oil phase is above CMC).

We formulated this hypothesis based on a series of initial experiments using brightfield optical microscopy in a sandwich-cell geometry depicted in Figure 2A. Movie S1 corresponds to the scenario depicted in Figure 1B of an oil-aqueous (at pH 3) interface. We note that there is no evidence of an interfacial instability from movie S1. Similarly, movie S2 corresponds the scenario depicted in Figure 1C of an oil-aqueous (at pH 7) interface- no interfacial flows were observed in this video either. Movie S3 corresponds to the scenario depicted in Figure 1D wherein 2 mM of $ZnSO_4$ is solvated in the pH 7 aqueous buffer. Interestingly, in movie S3 we see an interface that is unstable, and emulsions (that form spontaneously) enable the visualization of flow. The circulation patterns observed here are a classic signature of Marangoni rolls resulting from interfacial gradients [22, 55-58]. Flows in movie S3 were sustained for about three hours (HDEHP concentration in the oil phase was 50mM).

Interfacial instabilities were observed over a broad range of HDEHP concentrations within the organic phase, spanning 0.5–100 mM (concentrations above 100 mM were not explored, but it is expected that the instability would persist at those concentrations). Of particular significance, we do not observe interfacial instabilities below HDEHP concentration ~ 0.1 mM within the organic phase. The critical micellar concentration (CMC) of HDEHP in hexadecane is reported to be around~ 0.1-0.5 mM [49]. This observation is consistent with our suggested mechanism in Figure 1D, wherein, the removal of HDEHP-metal complex from the interface into the organic phase has to be accompanied with the ability to replenish the interface with fresh surfactant, which can only occur above the CMC [17, 18].

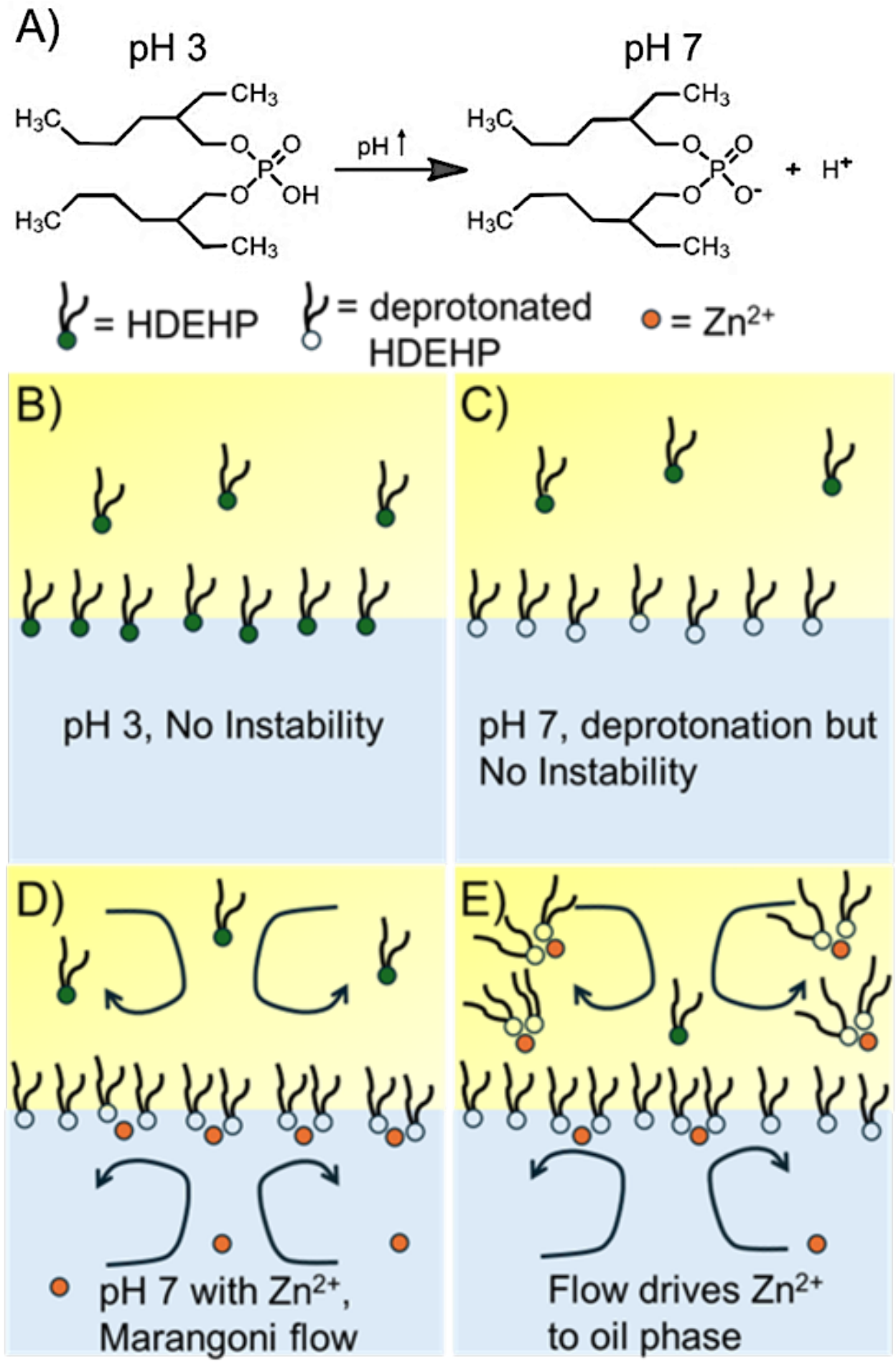

**Figure 1**. (A) Molecular structure and pH-dependent protonation states of HDEHP. Schematic illustration depicting oil-aqueous interface at different aqueous phase pH and salt content; (B) pH 3, (C) pH 7, (D) pH 7 with $Zn^{2+}$, and (E) pH 7 with $Zn^{2+}$ driven across the interface, respectively.

**2.2 Characterization of flow**: Using the geometry depicted in Figure 2A, direct observations of interfacial motion were made using time-resolved brightfield microscopy. Figure 2B is an image of the interface of hexadecane (with 50 mM HDEHP) and pH 3 aqueous buffer with 2 mM $ZnSO_4$ solvated within. Under these conditions, the interface remained quiescent over observation periods extending for several hours (Movie S4). In contrast, when the aqueous phase was maintained at pH 7, the interface exhibited pronounced spontaneous activity (Figure 2C and Movie S3). Under these conditions, persistent Marangoni flows were observed at the interface. Emulsions that spontaneously form at the interface help with the flow visualization and particle tracking to characterize the flow [12, 17, 23, 25, 48, 59-63]. Figure 2D shows the trajectory of an emulsion droplet entrained within a single Marangoni roll (Movie S5). The particle (2μm) completed an entire circulation loop within 0.67 s, velocity =550 μm $s^{-1}$ (laminar flow as Reynolds number ($\frac{\rho v L}{\mu}$) = 0.11). Velocity was also measured for emulsions (3μm) entrenched between the two Marangoni rolls (Figure 2E). Here, the average particle velocity increased from 69 μm $s^{-1}$ (± 0.88) at 100 µm from the interface to 142 µm $s^{-1}$ (± 1.35) directly close to the interface (Movie S6).

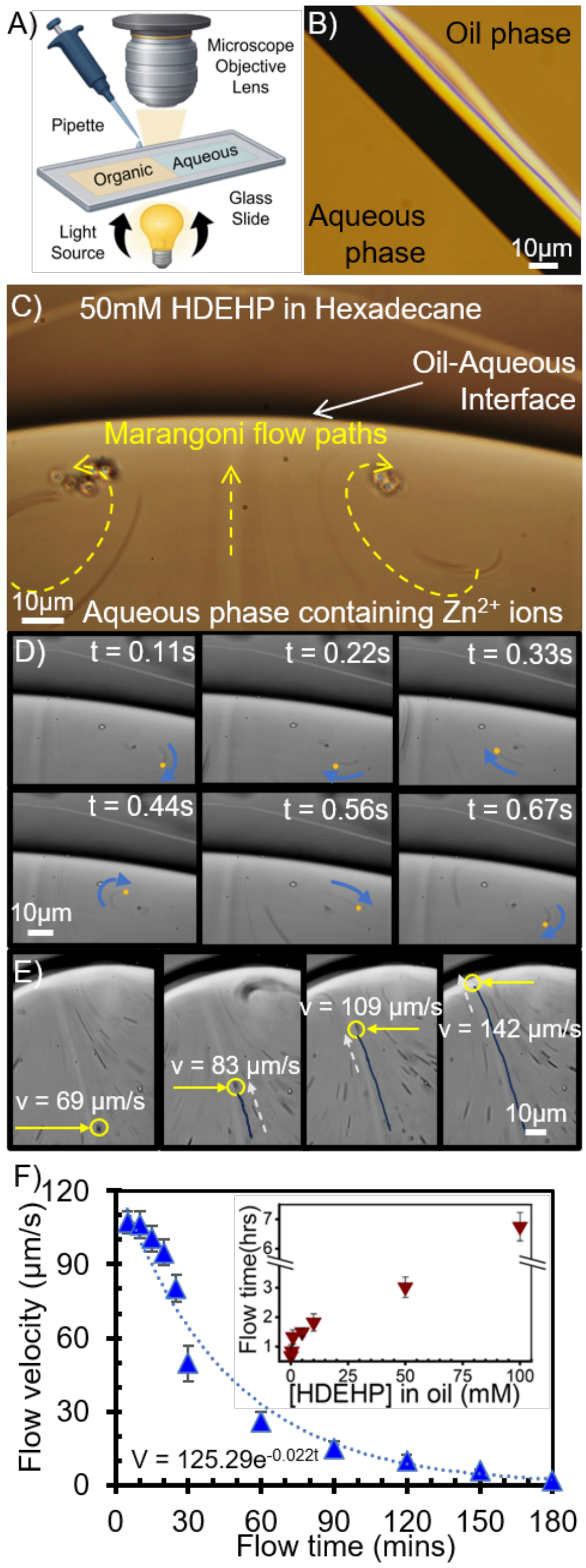

Figure 2. (A) Schematic of the microscopy setup, (B) Micrograph of oil-aqueous (pH~3) interface. No Marangoni rolls were observed (C) Micrograph of oil-aqueous interface showing the Marangoni rolls and path, aqueous phase pH is 7 (D) Snapshots showing particle movement within a Marangoni roll in the aqueous phase (E) Micrographs showing active transport of an emulsion towards the interface via the channel between the two rolls (F) Velocity profile of an emulsions passing through the channel between the rolls as a function of time. Inset in (F) Marangoni flow time versus HDEHP concentration in the oil phase. 50mM HDEHP in hexadecane (organic phase) and 2mM $ZnSO_4$ in phosphate buffer pH 7 (aqueous phase) were used for all the experiments in this figure.

Additionally, we measured the velocity of particles (~ 3 µm) at a distance~ 30 µm from the interface as a function of time (Figure 2F). Over a period of three hours the velocity at this location decreases from 105 µm $s^{-1}$ to nearly zero. The decrease in velocity follows an exponential decay, akin to other active droplet systems[25, 30], with a characteristic decay time of ~ 45 mins. The decrease in velocity of active droplets often follows an exponential decay as the driving force responsible for motion gradually weakens in a first-order relaxation process[24]. As these gradients dissipate through diffusion, reactant depletion or adsorption equilibration, the propulsion force decreases proportionally to its instantaneous magnitude thereby producing exponential behaviour [24, 25, 29, 30, 64]. The inset to Figure 2F characterizes the flow time as function of HDEHP concentration demonstrating that both the magnitude and lifetime of Marangoni activity depend strongly on HDEHP concentration. Specifically, the duration of sustained convection increased from approximately 40 min at 0.1 mM HDEHP to nearly 6.75 h at 100 mM HDEHP. The extended convection lifetime at higher HDEHP concentrations likely arises from a larger surfactant reservoir that continuously replenishes interfacial concentration gradients, thereby sustaining Marangoni stresses and delaying exhaustion of the driving force.

**2.3 Liquid-Liquid Extractions via Marangoni Flows**: Next, we sought to explore the question "are the interfacial flows coupled to the extraction of metal ions from the aqueous phase?". Figure 3A is an image of a simple bench-scale liquid-liquid extraction (LLE) system. The aqueous phase consisted of 2 mM $Zn^{2+}$ in phosphate buffer at pH 7, while the organic phase comprised of hexadecane containing HDEHP at concentrations of 0.01 mM (below CMC, vial on the left) and 10 mM (above CMC, vial on the right). As can be observed from Figure 3A, for the 0.01 mM HDEHP system, that the organic phase remained optically clear while for HDEHP concentration of 10 mM, organic phase became visibly turbid within ~ 20 mins. In addition, we also find via time-lapse imaging (videos are sped up 16X) that the oil-aqueous interface is relatively stable when HDEHP is below the CMC value (0.01 mM) as seen in Movie S7. Strikingly, with HDEPH concentration above the CMC, the oil-aqueous interface is clearly unstable, even to the naked eye, as can be evidenced in Movies S8 (10 mM HDEHP) and S9 (50 mM HDEHP). These two macroscopic observations provide hints that extractions under the presence of Marangoni flows are qualitatively distinct from diffusion-mediated transport.

To quantitatively understand the extraction behaviour with and without the presence of Marangoni flows, we employ Inductively Coupled Plasma Mass Spectroscopy (ICP-MS) to study the aqueous phase samples at various times. To show the generality of the phenomena, several divalent metals in addition to zinc were studied. Figures 3B and 3C plot the extraction kinetics for $Co^{2+}$ and $Ni^{2+}$ ions initially dissolved at

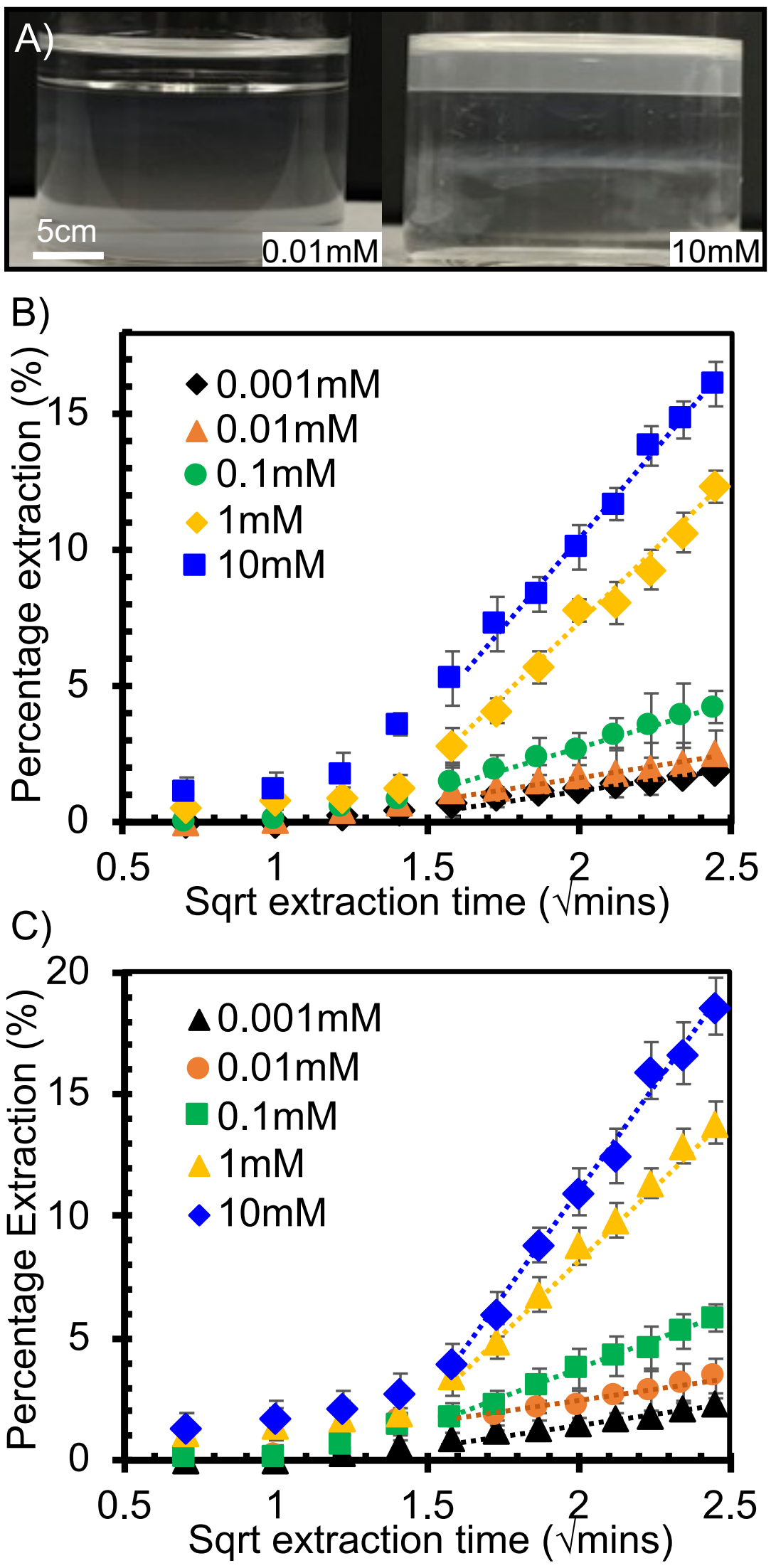

**Figure 3.** (A) Observations for a simple LLE set-up with the aqueous phase (2mM $ZnSO_4$ in phosphate buffer pH 7) interfaced with 0.01mM and 10mM oil phase concentration (HDEHP in hexadecane) respectively, (B) Percentage extraction of cobalt versus time for different HDEHP concentrations in the oil phase (C) Percentage extraction of Nickel versus time for different HDEHP concentrations in the oil phase. Dashed lines are fits to the long-timescale data.

2 mM in pH 7 aqueous buffer. Extraction kinetics for both metal ions show several key qualitative similarities. For concentrations of HDEHP below CMC (namely, .001 mM, .01 mM, .1 mM) within the oil phase, the temporal evolution of extraction percentage of the ions shows a linear dependence when plotted vs $t^{1/2}$. This is consistent with diffusion-mediated extraction behaviour. In contrast, the extraction profiles obtained at HDEHP concentrations above CMC (namely 1mM and 10 mM) deviated strongly from linearity. This departure from diffusive scaling indicates that the extraction process is likely coupled to the Marangoni flows and no longer follows diffusion-limited transport alone.

For the 1 mM and 10 mM data (> CMC), we notice that short-time data is non-linear, while the extraction percentages at long-timescales shows a linear behaviour with an significantly higher slope compared to data below CMC (analogous to an enhanced diffusion coefficient of active droplets after rotational diffusion randomizes the ballistic trajectory)[65]. In the classical two-film description of mass transfer processes [66-68], the central idea is that a fluid film or mass transfer boundary layer forms wherever there is contact between two phases. The thin film is treated to be relatively stagnant on either side of the interface and mass transfer through this film is affected solely by molecular diffusion. If this classical model applies to our system, our long-time data indicates that Marangoni flows, in effect, enhance the "net mass diffusion constants" and therefore enable efficient extractions. Comparing the slopes of the linear portions (dashed lines in Figure 3B and 3C, $\sqrt{t} = 1.58$ to 2.45) of the extraction profiles with those of the below CMC values reveals a 3-fold (or greater) increase in the extraction rates when Marangoni flows are present.

**2.4 Coupling between Interfacial Tension measurements and Extractions:** Since we posit that the metal-HDEHP binding at the oil-water interface is coupled to the generation of Marangoni flows, we sought to explore if differences in cation–surfactant interaction result in differential rates of transport for a range of divalent metals. Specifically, we sought to determine whether metal ions can produce differential interfacial tension gradients and in turn show varied extraction kinetics. To probe this possibility, the interfacial behaviour of four divalent transition-metal ions ($Cu^{2+}$, $Co^{2+}$, $Ni^{2+}$, $Zn^{2+}$) were examined systematically via pendant-drop tensiometry. Figure 4A is a schematic showing the silhouette of an axisymmetric fluid droplet in a pendant-drop tensiometry set-up. We rescale the bond number ($B_o \equiv \frac{\Delta\rho g R_o^2}{\gamma}$) in terms the Worthington number ($W_o = \frac{\Delta\rho g V_d}{\pi\gamma D_n}$)[69, 70] to account for correction due to needle diameters[69]. $\Delta\rho$ is the density difference between the aqueous droplet and ambient oil phase, $g$ is gravitational acceleration, $R_0$ is radius of the aqueous droplet, $V_d$ is the droplet volume, $\gamma$ is the interfacial tension, and $D_n$ is the needle diameter. To ensure reproducibility, each sample composition was measured at least 10 times, and only data with Worthington numbers~ 0.65-0.80 were used for measurements[69] of $\gamma$. We report the decrease in interfacial tension ($\Delta\gamma_0$) when an additional metal salt is present in phosphate buffer (with respect to the value of phosphate buffer and oil+ HDEHP interface). Figure 4B shows the $\Delta\gamma_0$ values for the four metal ions at a HDEHP concentration of 0.001 mM (well below the CMC). As we are well below the CMC, we posit that any differences in cation affinity can be captured via interfacial tension measurements (without the complication of interfacial convection). Among the ions examined, $Zn^{2+}$ produced the largest $\Delta\gamma_0$ (7.61 ± 0.39 mN/m), and this decrease was significantly greater than those measured for $Ni^{2+}$ (3.73 ± 0.34 mN/m), $Co^{2+}$ (3.72 ± 0.40 mN/m), or $Cu^{2+}$ (2.81 ± 0.41 mN/m). This trend in $\Delta\gamma_0$ values as a function of metal salt is maintained for a range of HDEHP concentrations below CMC

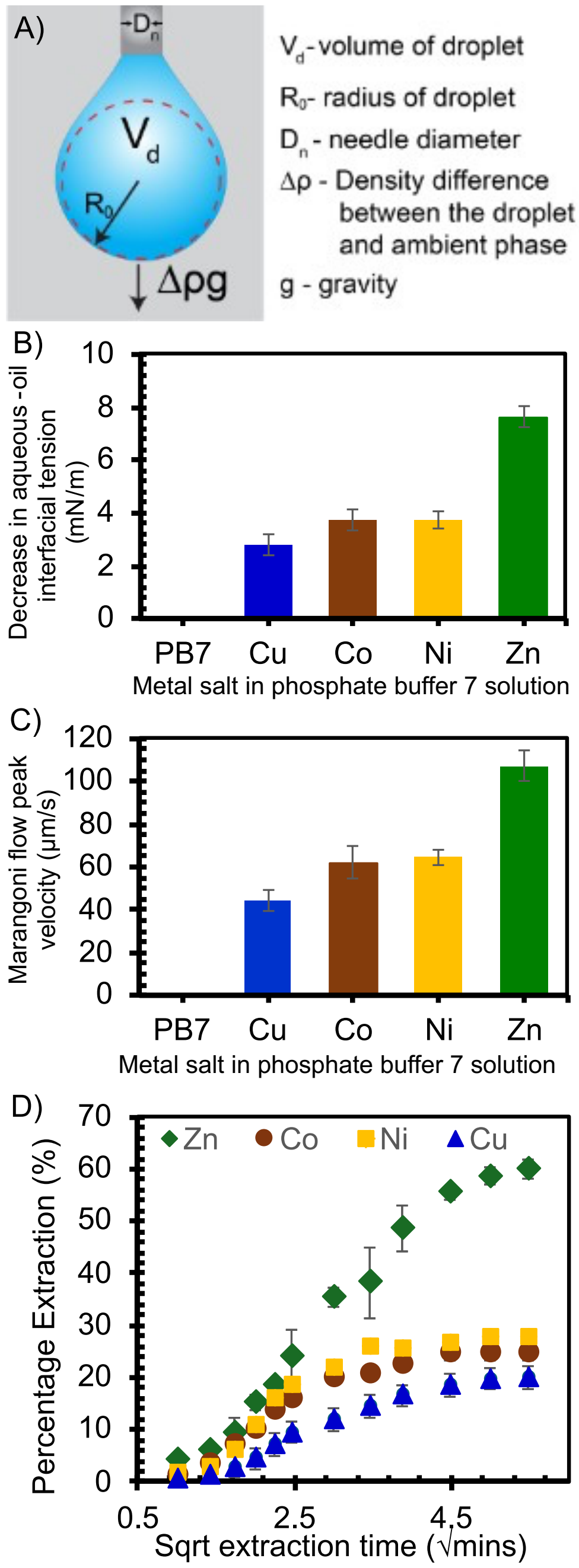

**Figure 4.** (A) Schematic of the pendant drop tensiometry setup for a water-in-oil droplet, (B) Aqueous-oil interfacial tension as a function of metal ions in aqueous buffer phase, (C) Marangoni flow peak velocity as a function of metal ions in aqueous buffer phase, (D) Percentage extraction versus time for various metal ions from independent aqueous metal salt solutions.

and interestingly also for 1 mM (above CMC) HDEHP- this is presented in Table S2.

We note that the measured flow velocities (Figure 4C) and the extraction profiles (Figure 4D) for these divalent metals are correlated to the decreases in their interfacial tension measurements. $Zn^{2+}$ generated the largest Marangoni velocity (107.17 ± 2.42 μm $s^{-1}$), followed by $Ni^{2+}$ (64.75 ± 1.14 μm $s^{-1}$), $Co^{2+}$ (62.17 ± 2.51 μm $s^{-1}$), and $Cu^{2+}$ (44.21 ± 1.57 μm $s^{-1}$). This close correspondence provides evidence that the intensity of the convective instability is coupled to the magnitude of the interfacial tension gradient established by each cation. We note that for a particle of radius $R$ moving at velocity $v$, interfacial dissipation scales as $Rv\Delta\gamma$, where $\Delta\gamma$ is the instantaneous gradient of the interfacial tension along the interface (Figure 1D and 1E). Equating this to viscous dissipation $\mu R v^2$ results in the classical definition of capillary number as follows:

$$C_a = \frac{\mu v}{\gamma} \approx \frac{\Delta\gamma}{\gamma} \qquad (1)$$

where μ is the shear viscosity.

From equation 1, the Marangoni velocity can be expressed as:

$$v \approx \frac{\Delta\gamma}{\mu} \qquad (2)$$

As typical viscosities at oil-water interfaces are of the order ~ 1 mPa·s [25, 29, 61, 71], simple scaling analysis reveals that for the dynamic oil-water interface in our system, $\Delta\gamma$, is in the range of ~ $10^{-4}$ mN/m. These low values of $\Delta\gamma$ have been predicted for other active droplets systems previously[19]. The observation of $\Delta\gamma << \Delta\gamma_0$ is unsurprising as the presence of Marangoni flows dynamically rearrange the interface and therefore the gradients are expected to be lower than $\Delta\gamma_0$ which is an equilibrium scenario with absence of any interfacial flows.

To assess whether these extraction rates can be attributed/coupled to interfacial tension differences, independent experiments were performed using 2 mM $Cu^{2+}$, $Co^{2+}$, $Ni^{2+}$, and $Zn^{2+}$ solvated (sulphate counterion) in phosphate buffer solutions at pH 7. The aqueous phase was contacted with 10 mM HDEHP in the organic phase for 30 min. The resulting extraction profiles are shown in Figure 4D. In addition to the short-time and long-time behaviour that is reminiscent of the data presented in Figure 3, we also note that the extraction for most metals (except zinc) saturates at around t~ (20 mins). From Figure 4, we note that the extraction trends matched interfacial tension and velocity trends. $Zn^{2+}$ showed the largest $\Delta\gamma_0$, highest Marangoni velocity, and greatest extraction rates. $Ni^{2+}$ and $Co^{2+}$ showed intermediate extraction rates, whereas $Cu^{2+}$ exhibited the lowest extraction efficiency. This agreement between interfacial tension and extraction behaviour highlights the key role of ion-extractant affinity in governing interfacial transport dynamics. These findings demonstrate that metal-dependent changes in the interfacial free-energy landscape are coupled to extraction profiles under non-equilibrium conditions.

## 3. Conclusions

This work establishes a non-equilibrium liquid-liquid extraction framework, showing that Marangoni flows can be harnessed to enhance metal ion transport across oil-water interfaces. By coupling pH-dependent extractant chemistry with ion-specific adsorption, extraction profiles show a qualitative shift from diffusion-limited to convection-dominated transport.

Deprotonation of HDEHP and metal-ion adsorption generate interfacial tension gradients that drive sustained Marangoni convection, enhancing mass transfer ~3-fold over equilibrium extraction rates. Time-resolved optical microscopy reveals that these flows occur only when three conditions are met: 1) pH is maintained well above the extractant pKa, 2) interfacially active metal cations present, 3) surfactant concentration above a critical threshold- closely coupled to its CMC. The lifetime and intensity of Marangoni activity increase with surfactant concentration, providing a kinetic handle for tuning interfacial transport. Pendant-drop tensiometry shows metal-dependent interfacial tension reductions that correlate with extraction rates; $Zn^{2+}$ induces the largest decrease and is preferentially extracted over $Ni^{2+}$, $Co^{2+}$, and $Cu^{2+}$ under Marangoni-driven conditions. Importantly, this work hints that selective extractions are possible as extraction rates are coupled to the binding affinities of metal-extractant.

Our work also raises several open questions that remain to be answered. We are yet to understand the origin of the interfacial gradients and their coupling to the molecular details of the extraction mechanism, for instance, the nature of the complexes formed at the interface, their desorption rates etc. It also remains unclear whether the observed spontaneous emulsification is relevant to the metal transport across the interface. Effect of valency of the metal ion on the extraction kinetics remains to be studied.

Overall, these findings open a new avenue for integrating active-matter concepts with liquid-liquid extraction phenomena with the potential to enable faster, selective, and solvent-efficient separations.

## 4. Conflicts of interest

No conflicts to declare.

## 5. Acknowledgements

KN acknowledges funding from American Chemical Society Petroleum Research Fund.

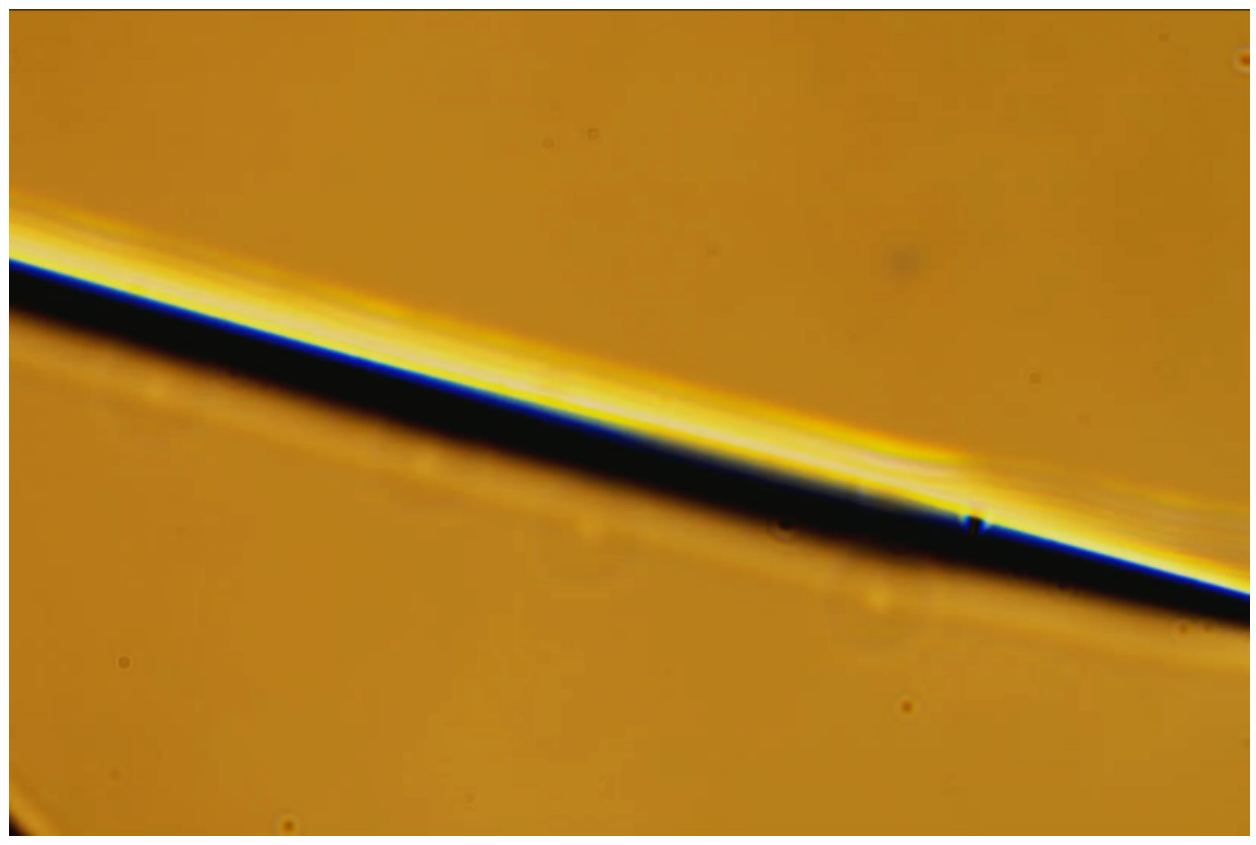

Movie S1. Oil-aqueous interface with the aqueous phase at pH 3 and no metal salt solvated within. Oil phase contained 50mM HDEHP in hexadecane. No instabilities (Marangoni rolls/path) were observed in this video.

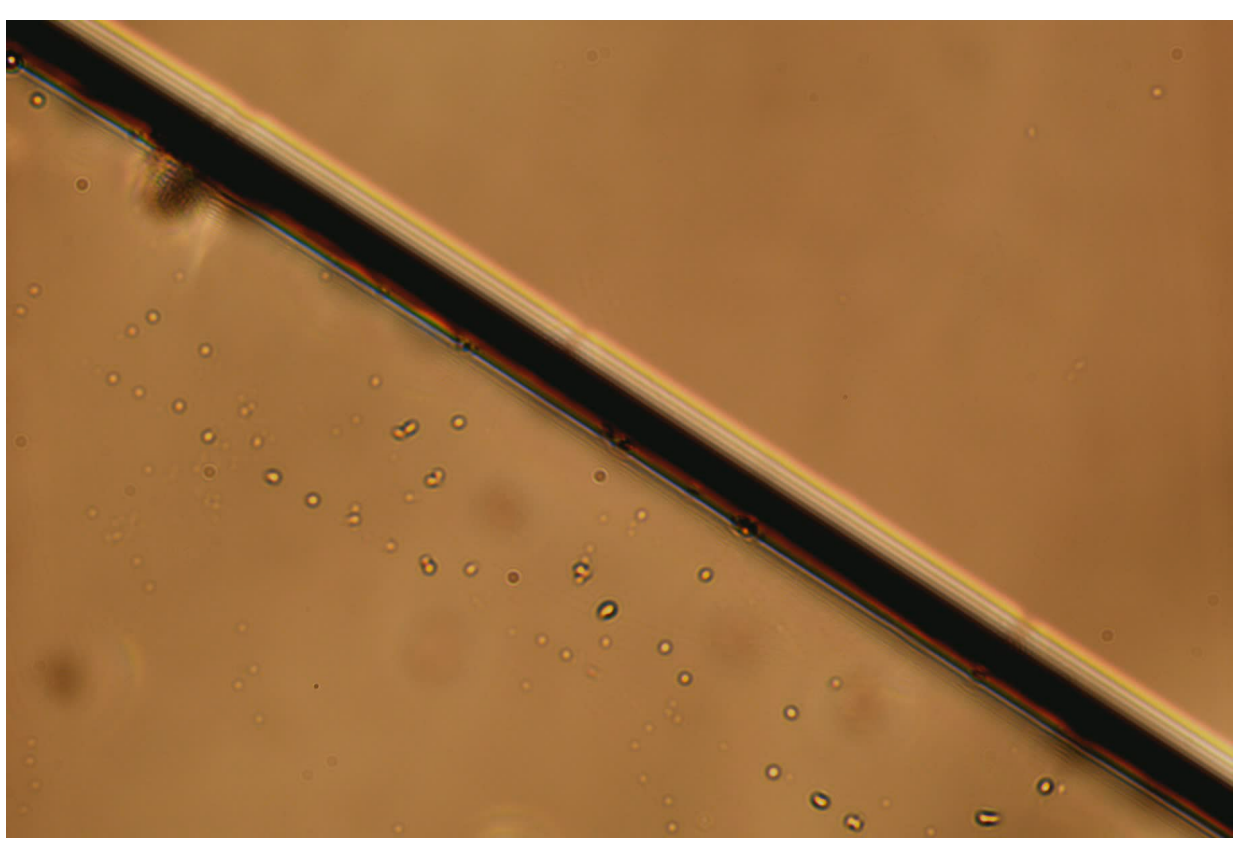

Movie S2. Oil-aqueous interface with the aqueous phase at pH 7 and no metal salt solvated within. Oil phase contained 50mM HDEHP in hexadecane. No instabilities (Marangoni rolls/path) were observed in this video.

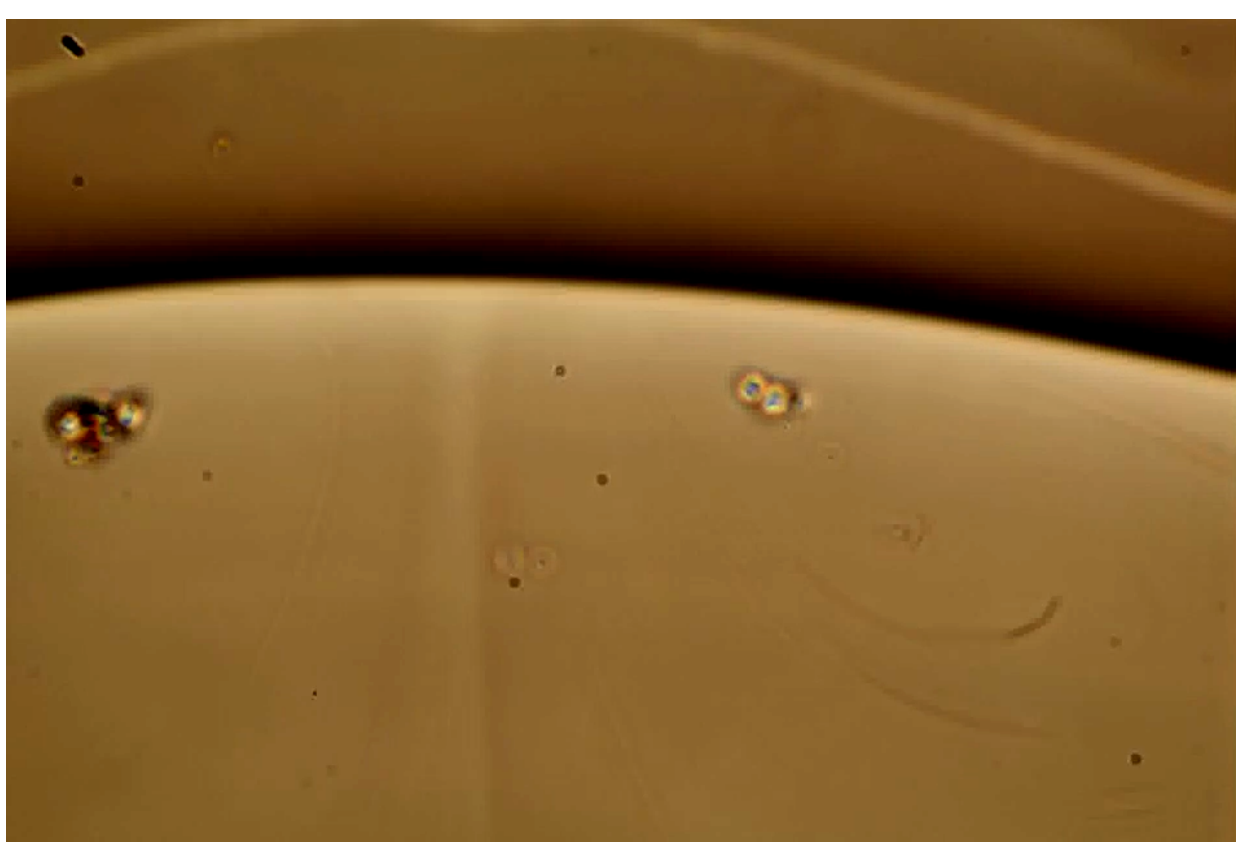

Movie S3. Oil-aqueous interface with the aqueous phase at pH 7 and 2mM $ZnSO_4$ solvated within. Oil phase contained 50mM HDEHP in hexadecane. Instabilities (Marangoni rolls/path) were clearly observed for this setup.

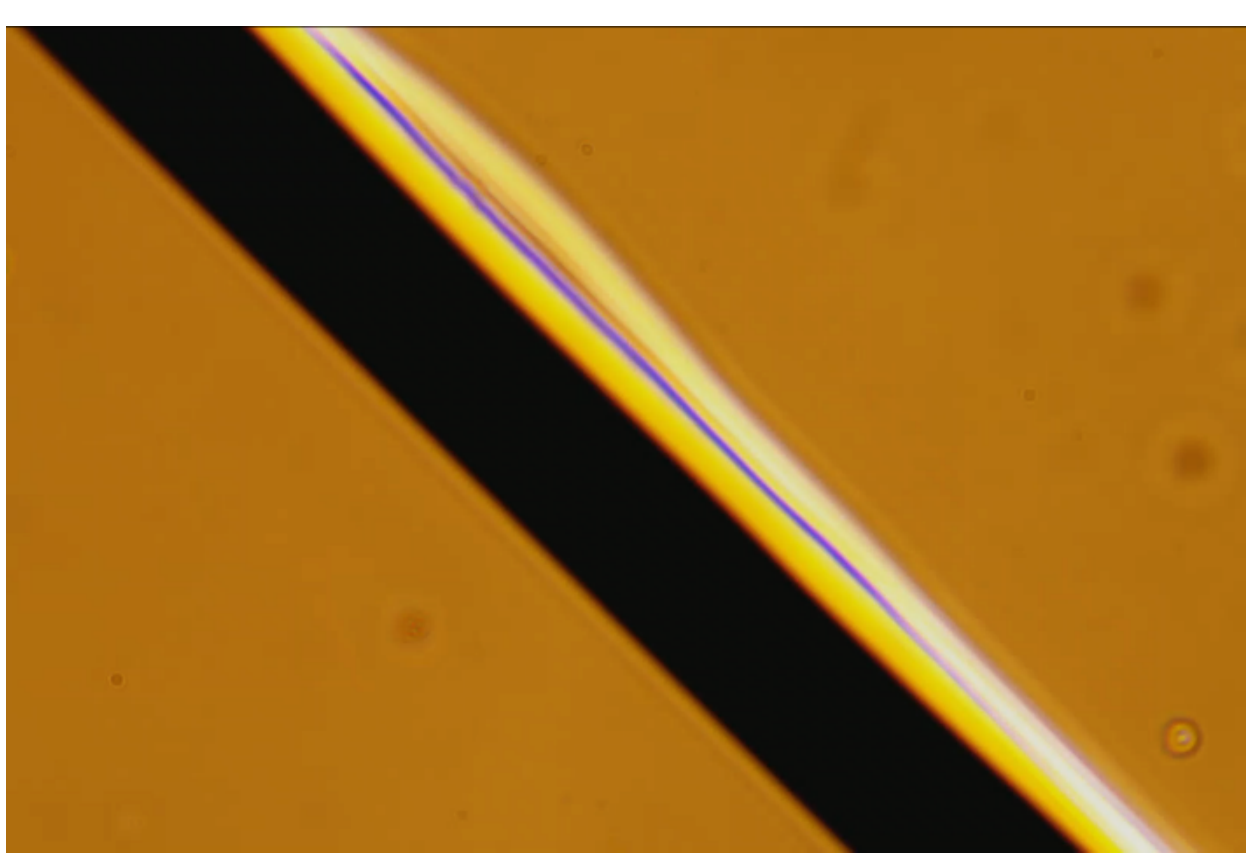

Movie S4. Oil-aqueous interface with the aqueous phase at pH 3 and 2mM $ZnSO_4$ solvated within. Oil phase contained 50mM HDEHP in hexadecane. No instabilities (Marangoni rolls/path) were observed in this video.

Table S1. Marangoni flow time as a function of HDEHP concentration

| HDEHP Concentration in hexadecane (mM) | Flow time (hrs) | Flow type |
|---|---|---|
| 0.01 | - | No Marangoni flow |
| 0.05 | - | No Marangoni flow |
| 0.1 | 0.67 | Weak Marangoni flow |
| 0.5 | 0.83 | Weak Marangoni flow |
| 1 | 1.33 | Marangoni flow |
| 5 | 1.50 | Marangoni flow |
| 10 | 1.83 | Marangoni flow |
| 50 | 3.00 | Marangoni flow |
| 100 | 6.75 | Marangoni flow |

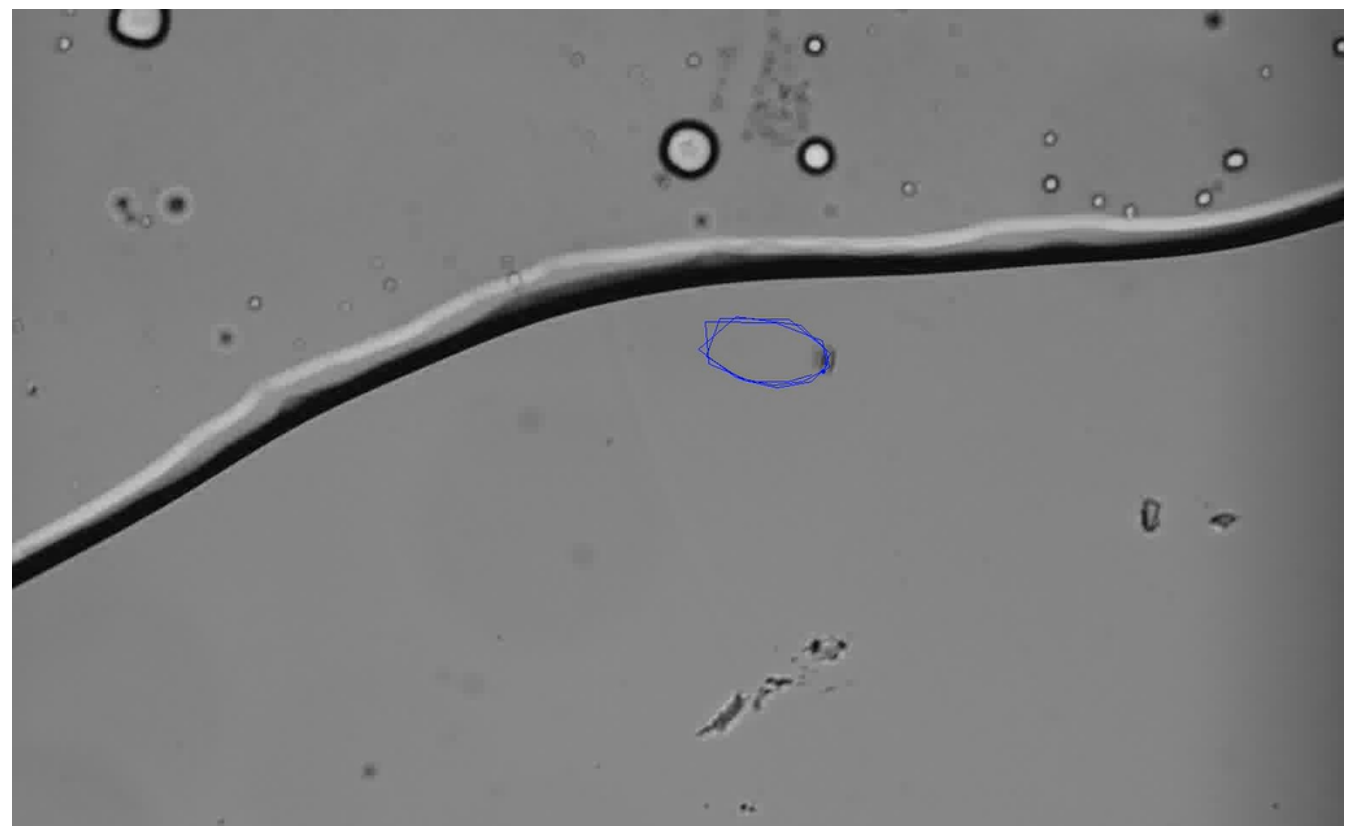

Movie S5. This video shows the trajectory of an emulsion particle entrained within a single Marangoni roll. Oil phase contained 50mM HDEHP in hexadecane and aqueous phase contained metal salt (2mM $ZnSO_4$) solvated within.

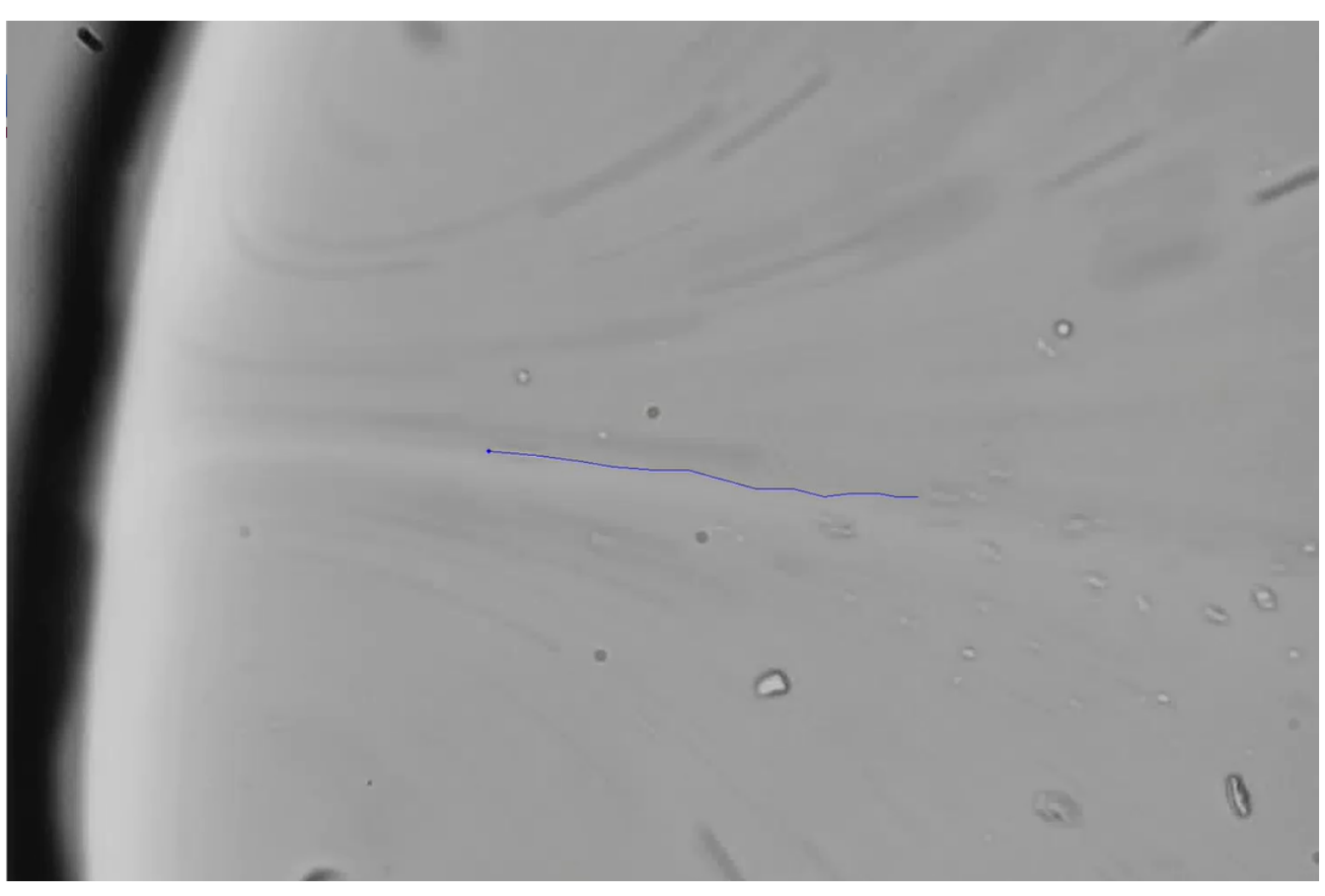

Movie S6. This video shows the trajectory of an aqueous-phase particulate entrained in the path between two Marangoni rolls. Oil phase contained 50mM HDEHP in hexadecane and aqueous phase is a pH 7 buffer containing 2mM $ZnSO_4$.

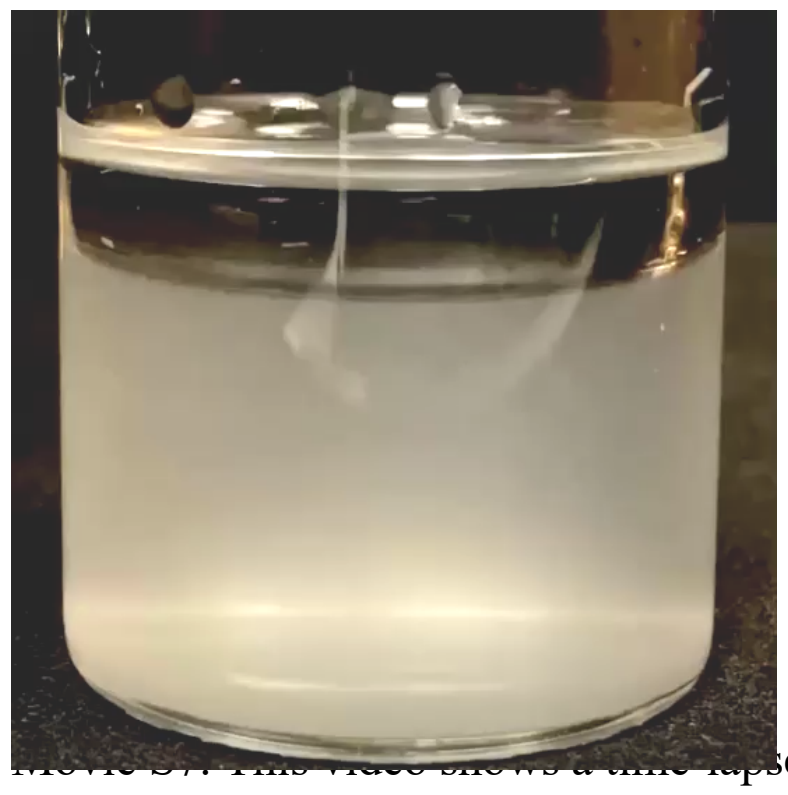

Movie S7. This video shows a time-lapse of lab-scale liquid-liquid extraction setup, with 0.01mM HDEHP in hexadecane (organic phase, top) and 2mM $ZnSO_4$ in phosphate buffer pH 7 (aqueous phase, bottom).

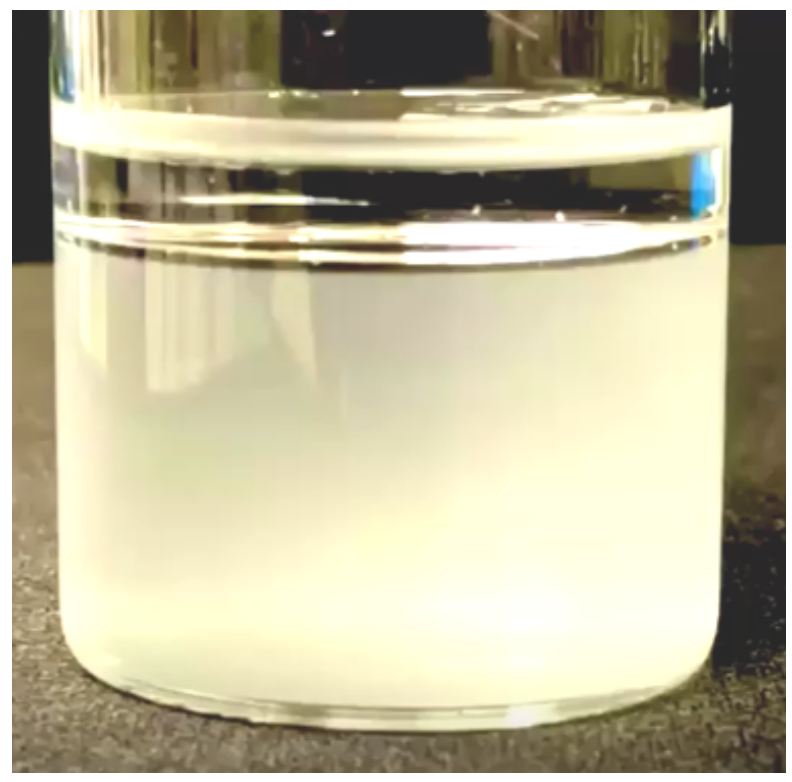

Movie S8. This video shows a time-lapse of lab-scale liquid-liquid extraction setup, with 10mM HDEHP in hexadecane (organic phase, top) and 2mM $ZnSO_4$ in phosphate buffer pH 7 (aqueous phase, bottom).

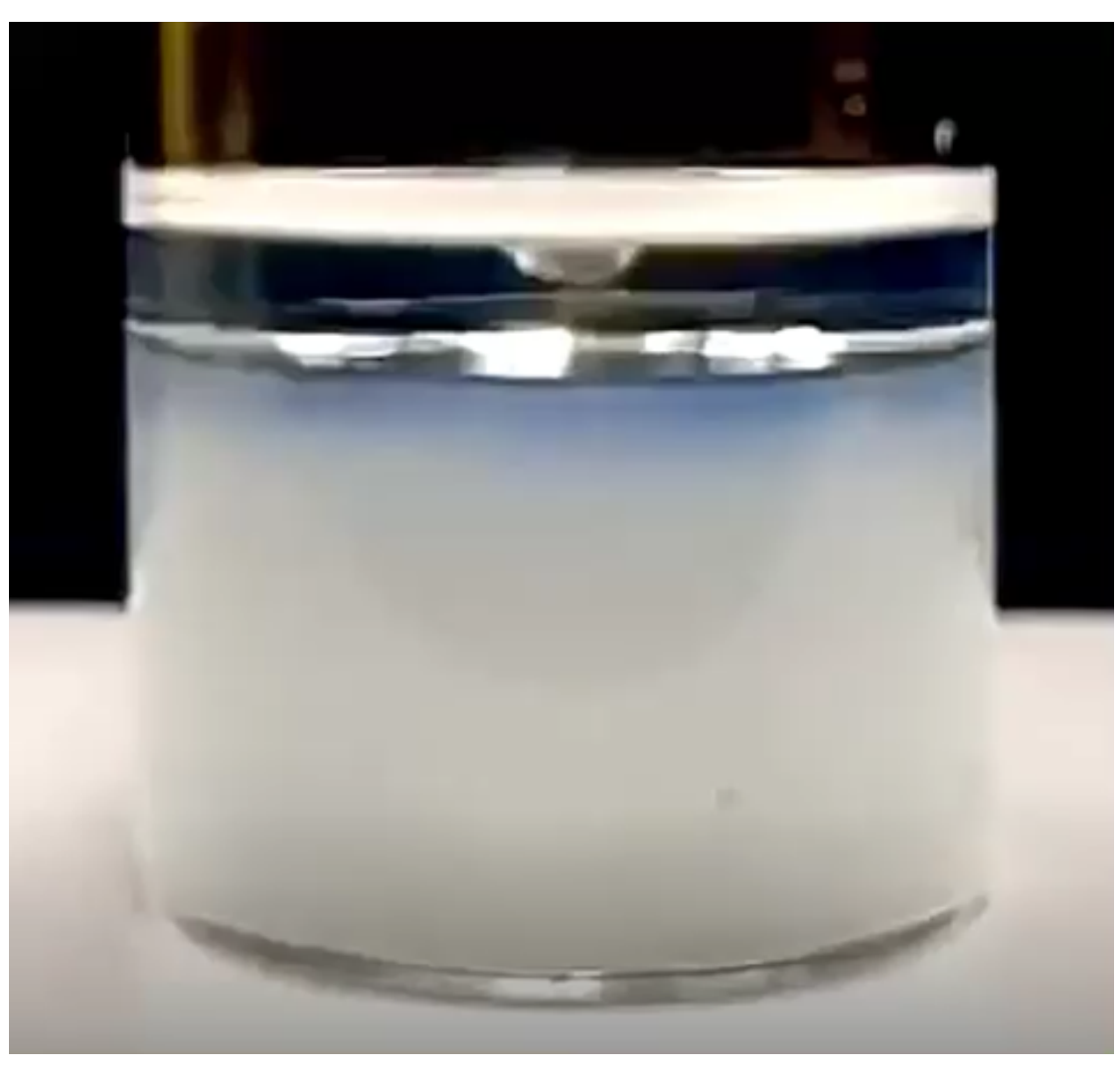

Movie S9. This video shows a time-lapse of lab-scale liquid-liquid extraction setup, with 50mM HDEHP in hexadecane (organic phase, top) and 2mM $ZnSO_4$ in phosphate buffer pH 7 (aqueous phase, bottom).

Table S2. Decrease in aqueous-oil interfacial tension (mN/m) when metal salts are added to the aqueous phase, with respect to phosphate buffer 7 (control)

| **Test Item (Metal salt in phosphate buffer 7)** | **Decrease in aqueous-oil interfacial tension (IFT) across varying HDEHP concentrations (mN/m)** | | | |
|---|---|---|---|---|
| | **0.001 mM** | **0.01 mM** | **0.1 mM** | **1 mM** |
| PB7 | 0 | 0 | 0 | 0 |
| Cu | 2.8068 | 2 | 1.209 | 0.2472 |
| Co | 3.7199 | 1.529 | 2.6178 | 0.5072 |
| Ni | 3.7389 | 1.2619 | 3.1583 | 0.8204 |
| Zn | 7.618 | 2.5107 | 3.6314 | 1.0983 |
| La | 8.6947 | 3.1677 | 5.8322 | 1.1802 |